# OMG-VR: Open-source Mudra Gloves for Manipulating Molecular Simulations in VR

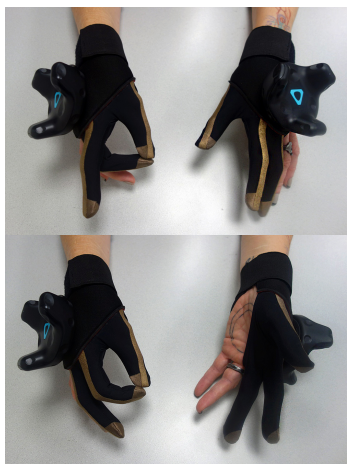

Figure 1: Our application specific pinch sensing gloves (above) are adapted from a previous open source design (below).

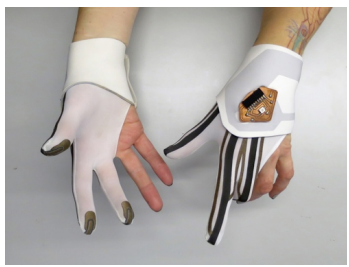

**Rachel Freire**
Rachel Freire Studio & Intangible Realities Laboratory
London/Bristol, UK
rachel@rachelfreire.com

**Becca Rose Glowacki**
Department of Design
Goldsmiths College
University of London
b.glowacki@gold.ac.uk

**Rhoslyn Roebuck Williams**
Intangible Realities Laboratory
University of Bristol
Bristol, UK
rhoslyn.roebuckwilliams@bristol.ac.uk

**Mark Wonnacott**
Intangible Realities Laboratory
University of Bristol
Bristol, UK
m.wonnacott@bristol.ac.uk

**Alexander Jamieson-Binnie**
Intangible Realities Laboratory
University of Bristol
Bristol, UK
alexander.jamieson-binnie@bristol.ac.uk

**David R Glowacki**
Intangible Realities Laboratory
University of Bristol
Bristol, UK
glowacki@bristol.ac.uk

## Abstract
As VR finds increasing application in scientific research domains like nanotechnology and biochemistry, we are beginning to better understand the domains in which it brings the most benefit, as well as the gestures and form factors that are most useful for specific applications. Here we describe Open-source Mudra Gloves for VR (OMG-VR): etextile gloves designed to facilitate research scientists and students carrying out detailed and complex manipulation of simulated 3d molecular objects in VR. The OMG-VR is designed to sense when a user pinches together their thumb and index finger, or thumb and middle finger, forming a "mudra" position. Tests show that they provide good positional tracking of the point at which a pinch takes place, require no calibration, and are sufficiently accurate and robust to enable scientists to accomplish a range of tasks that involve complex spatial manipulation of molecules. The open source design offers a promising alternative to existing controllers, and more costly commercial VR data gloves.

## CSS Concepts
• Human-centered computing ~ Virtual reality

## Introduction: Molecules in Virtual Reality
Nanoscale molecular objects offer fertile testbeds for applying new methods in human-computer interaction,

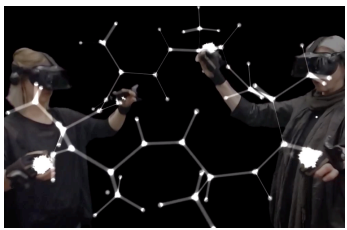

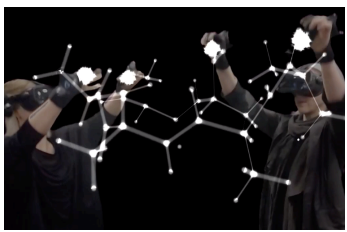

Figure 2: People wearing the open data gloves to manipulate a real-time simulation of a poly-alanine protein (images taken from corresponding video vimeo.com/356981043)

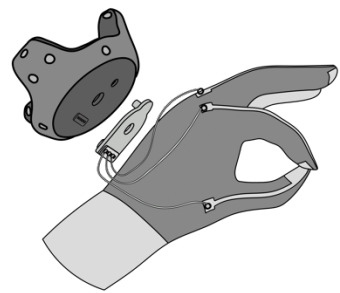

Figure 3: HTC vive tracker attaches to 3d printed mount that houses pogo pins. Pins connect to the silver plated fabric.

owing to the fact that molecules represent objects that are important to society and industry, but which we are unable to directly perceive, and which are characterized by considerable three-dimensional dynamic complexity that can be accurately simulated. Over the past few years we have developed an open-source VR framework called Narupa [10], which enables multiple participants to co-habit a virtual environment and interact with complex molecular structures whose non-intuitive dynamics and physics are simulated in real-time. Narupa is designed using a client/server architecture, in which each room-scale VR client (e.g., HTC Vive, Valve Index, etc.) connects to a real-time molecular physics server. To date, Narupa has found application as a tool for nanotechnology research [15] and education [1]. In a recent study, we examined how quickly scientists could accomplish different molecular research tasks using VR versus standard 2d interfaces [14]. For tasks requiring complex 3d spatial manipulation, we found that VR offered significant accelerations: e.g., when we asked scientists to create knots in small protein molecule, we observed accelerations of ~10x.

## Accurately Sensing Virtual Objects

Several Narupa users have reported that it enables them to 'sense' the flexibility of different molecular physics simulations [15]. On a few occasions, they have commented that plastic handheld VR controllers limited their ability to accurately sense the mechanical properties of the simulated molecular objects. Such reports have inspired us to investigate mechanisms for enabling users to manipulate simulated molecular objects, without mediation by a controller [16]. From a scientific research and educational perspective, it is advantageous to design VR frameworks that researchers and students can use to develop an accurate sense of a molecule's dynamics and flexibility: it enables researchers to accomplish simulation tasks faster than they would otherwise, and it enhances the insight that both researchers and students can gain into how molecules behave, allowing them to make more intelligent hypotheses as to their design and function [2].

As an alternative to controllers, we have begun to explore the use of gloves, building on previous work where gloves have been used to enable people to "touch" virtual objects [8, 13]. As a first test of gloves in facilitating typical Narupa workflows, we interfaced the Narupa framework to two different commercial VR data gloves: the Manus [9] and the Noitom Hi5 [11]. These gloves are designed primarily for tracking of hand poses, to distinguish amongst a variety of gestures (e.g., between a 'thumbs-up' or a Vulcan 'live long and prosper' gesture) for application in environments like motion capture studios. Each is equipped with a 9-degree of freedom, inertial movement unit, with bend sensors mounted along the back of each finger. Both came equipped with a wrist mount to which an HTC Vive tracker can be attached, enabling absolute positional tracking by the HTC Vive lighthouse tracking system. We carried out extensive tests evaluating the ability of each of these glove models to facilitate efficient molecular manipulation and found their performance not particularly well suited to our purposes, for the following reasons: (1) Owing to the number of sensors in these gloves, they require frequent calibration to account for sensor drift and changes in position of the wrist mounted tracker. For example, calibration is required before and sometimes during a glove session, taking time and thus introducing a barrier to usage; (2) Neither glove is particularly reliable in detecting when a researcher is 'pinching' a *particular* atom (or selection of atoms), an important gesture for the molecular manipulation tasks that we regularly carry out with Narupa; (3) they are expensive (e.g., in Jan 2020, the Noitom Hi5 VR gloves cost $1k and the Manus VR gloves cost €3k, not including trackers); and (4) because they are size specific, they cannot be easily swapped between users. Given these issues, university laboratories like our own, which utilize the multi-person version of Narupa for both research and teaching, are unlikely to make such an outlay for non-commodity equipment, especially given that the Narupa VR framework is often used with 3–4 users, all of whom require a separate pair of gloves, with different hand sizes.

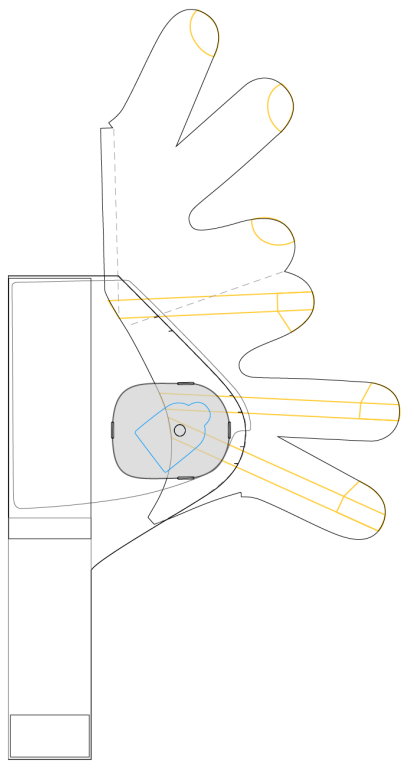

Figure 4: Sewing pattern for the V2.1 glove.

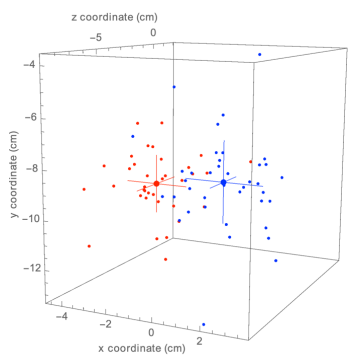

Figure 5: Pinch positions of left (blue) and right (red) hands of participants wearing OMG-VR, showing respective centroids and variances.

## An Open Source VR Glove Design

As an alternative to the commercially available gloves, we took inspiration from recent open-source data glove designs [3, 7, 12] to design low cost gloves (ranging between $50-100 not including trackers). The OMG-VRs are designed so that they can be made in-house and used for VR-enabled molecular manipulation by multiple glove-wearing participants (as shown in Fig 2 and the corresponding video at vimeo.com/356981043). The gloves (schematized in Fig 3) utilize a silver plated fabric circuit with pogo pins that connect to an HTC Vive tracker mounted on the back of the hand. The silver plated fabric is a stretch knit and the glove textile is a four-way stretch fabric. The circuit is bonded to the textile with a domestic iron using a heat-activated stretch adhesive film, reducing the amount of sewing required, and making the process more accessible to those with limited experience of working with fabric. This enables any given glove to fit a range of hand shapes, increases comfort for the user, and provides a larger error tolerance for the glove maker. Fig 3 shows the 3d printed connector that attaches to the glove and makes contact with the Tracker input pins. The silver plated fabric is used to construct two circuits that are completed when the thumb contacts either the index or the middle finger, forming a "mudra" position. Within Narupa, completing the thumb-index finger circuit exerts an atomic force, pulling the nearest atom towards the approximate point at which the pinch occurs, as shown in vimeo.com/356981043. When the thumb-middle finger circuit is completed simultaneously for both the left and right hand, the user can scale the size of the simulation, or rotate its orientation.

The OMG-VR design achieves absolute positional tracking by mounting an HTC Vive tracker to the back of the hand. The rationale for this design arises from the fact that the spatial location of a pinch is relatively invariant to the back of the hand. Fig 5 shows results from 13 different scientists in our lab (3 pinches each), using a coordinate axis whose origin we defined to be coincident with the center of the tracker's flat side (shown in Fig 3). Fig 5 shows that the respective x, y, z variances of the centroid pinch locations of right and left hands were ±1.7 (±1.4) cm, ±1.3 (±1.6) cm, and ±1.7 (±1.2) cm. Given these relatively small variances, we found that it is a reasonable approximation to simply locate the left and right hand pinch point at the respective left and right hand centroids in Fig 5. Compared to the wrist mounted trackers used in the commercial gloves we tested, our back-of-the-hand mount does not require calibration prior to a VR glove session. Because OMG-VR avoids bend sensors in favor of direct binary inputs for each pinch, they deliver little unnecessary data and require no extraneous hardware. As a result, we have found this design to be extremely reliable for molecular manipulation compared to the commercial gloves, owing to the fact that there are very few moving parts, and consequently few points of failure. This simplicity, along with detailed online instructions (instructables.com/id/Etextile-VR-Gloves-for-Vive-Tracker) [4], allowed us to develop several pairs of data gloves "in-house" with basic technical infrastructure of the sort that is found in accessible makerspaces: sewing machines, domestic and soldering irons, 3d printers, as well as conductive and non-conductive fabrics.

During field tests in Aug 2019, the gloves were used to facilitate a multi-person VR experience called "Is-ness": over 3 days, 4 pairs of gloves were worn by more than 64 people, for 45-60 mins at a time [6]. We observed zero failures, providing good evidence for the reliability of this design. Between sessions, we swapped gloves between different users, and observed that they could be intuitively utilized to successfully carry out molecular manipulation tasks without any need to re-calibrate, even in cases where the users had different sized hands. In the latest iteration of OMG-VR, we added enhanced stabilization to the tracker mount, further improving its performance (Fig 4). We published the first version of the OMG-VR in Jan 2019, showing that a back-of-the-hand mounted HTC Vive tracker offered significant advantages compared to commercial wrist-mounted designs [5]. By mid-2019, Manus VR adopted a similar design in their commercial VR glove, affirming the reliability of our open-source design.

## Conclusion

Our tests to date suggest that the OMG-VR design is effective for undertaking complex spatial molecular manipulation tasks in VR. Compared to commercial gloves we trialed, complex tasks like molecular knotting (vimeo.com/305823646) are considerably smoother and less error prone using our glove design. They require less hardware, no calibration, are much cheaper, and one size fits a wide range of hand sizes. In terms of user comfort, the silver plated Etextile material feels lighter and less cumbersome than the thicker commercial material. Positioning the tracker on the back of the hand minimizes error in hand tracking, enabling more specificity of hand angle and location, and smoother hand motion. Combined, these create a more embodied sense of molecular manipulation compared to the hand-held room scale VR controllers. The open-source design is supported by comprehensive online documentation, enableing makers to understand how the glove works and how to combine different construction options. This enables makers to build a glove tailored to their skill-set, choose materials that are more readily available, and adapt and develop their gloves to make iterative improvements for the benefit of the open source community, offering a robust and viable application specific alternative to the expensive commercial models.

Moving forward, we plan to carry out more studies to evaluate the extent to which the OMG-VR enables research scientists to carry out complex 3d molecular manipulation tasks like protein-drug binding, and also investigate their utility in other scientific simulation and visualization domains beyond molecular science. We plan to carry out detailed comparisons evaluating the extent to which glove models like these enable researchers to better sense of the systems with which they are interacting. Developing in-house skills though the open source approach makes maintenance and repair easier. We are currently developing tutorials showing alternate construction, repair methods, and a "no-sew" version, constructed by modifying pre-constructed gloves, and testing with our user community.